\documentclass[%
 aip,
 amsmath,amssymb,
 reprint,%
]{revtex4-1}

\usepackage{graphicx}
\usepackage{dcolumn}
\usepackage{bm}

\usepackage[utf8]{inputenc}
\usepackage[T1]{fontenc}
\usepackage{mathptmx}
\usepackage{natbib}
\usepackage{color}
\usepackage[ruled,linesnumbered]{algorithm2e}
\usepackage{xspace}

\begin{document}

\preprint{AIP/123-QED}

\title[Sample title]{Constructions and properties of a class of random scale-free networks}

\author{Xiaomin Wang}
\email{wmxwm0616@163.com}
\affiliation{%
Key Laboratory of High-Confidence Software Technology, Peking University, Beijing 100871, China
}%
\affiliation{ 
School of Electronics Engineering and Computer Science, Peking University, Beijing 100871, China
}%
\author{Fei Ma}%
\email{mafei123987@163.com}
%
\affiliation{ School of Electronics Engineering and Computer Science, Peking University, Beijing 100871, China
}


\date{\today}

\begin{abstract}
Complex networks have abundant and extensive applications in real life. Recently, researchers have proposed a large variety of complex networks, in which some are deterministic and others are random.  The goal of this paper is to generate a class of random scale-free networks. To achieve this, we introduce three types of operations, i.e., rectangle operation, diamond operation, and triangle operation, and provide the concrete process for generating random scale-free networks $N(p,q,r,t)$, where probability parameters $p,q,r$ hold on $p+q+r=1$ with $0\leq p,q,r\leq 1$.  We then discuss their topological properties, such as average degree, degree distribution, diameter, and clustering coefficient. Firstly, we calculate the average degree of each member and discover that each member is a sparse graph. Secondly,  by computing the degree distribution of our network $N(p,q,r,t)$, we find that degree distribution obeys the power-law distribution, which implies that each member is scale-free.  Next, according to our analysis of the diameter of our network $N(p,q,r,t)$, we reveal the fact that the diameter may abruptly transform from small to large. Afterwards, we give the calculation process of clustering coefficient and discover that its value is mainly determined by $r$. 
\end{abstract}
 
\maketitle

\begin{quotation}

It is well-known that complex networks, especially random scale-free networks, are ubiquitous in the world and are applied in many areas of scientific researches. In order to better understand random scale-free networks, this paper aims to generate a class of random scale-free networks by graphic operations based on probabilistic behaviors.  Our scale-free networks form a network space $\mathcal{N}(p,q,r,t)$ with the probability parameters $p,q$ and $r$ holding on $p+q+r=1$ with $0\leq p,q,r\leq 1$ at each time step $t$.  Each member $N(p,q,r,t)$ of the space $\mathcal{N}(p,q,r,t)$ is a dynamic network that will be constructed from $N(p,q,r,t-1)$ by three graphic operations, i.e., rectangle operation, diamond operation, and triangle operation, at time step $t$. We then show the topological structures of each network $N(p,q,r,t)$ by its average degree, degree distribution, diameter, and clustering coefficient. We discover that degree distributions of our networks $N(p,q,r,t)$ follow power-law distribution, which indicates that each network is scale-free. More importantly, our scale-free networks differ from the existing scale-free networks determined by only a single operation rule because we have considered three types of operations. Our network is suitable for capturing the nature of multiple operations. 
\end{quotation}

\section{\label{sec:level1}Introduction}

Over the course of recent decades, complex networks have attracted thousands of scholars from different fields and provided a mathematical tool that connects the real world with theoretical research. Therefore, complex networks were applied across a multitude of disciplines ranging from natural and physical sciences to social sciences and humanities, for instance, Internet\cite{Watts-1998}, World Wide Web (WWW)\cite{Albert-1999-1}, coauthorship network\cite{zhou-li-2017}, citation network\cite{Golosovsky-2017}, annotated network\cite{Newman-2016}, musical solos network\cite{Ferretti-2017}, protein network\cite{Sun-Quan2017}, information network\cite{Hou-mao2016}, peer-to-peer network\cite{Li-2016}, social network\cite{Zoller-2014}, immune network\cite{Agliari-2013}, pseudofractal scale-free web\cite{Peng-2015}, manage network\cite{Pandey-2019},  and so forth.

In general, a network can be viewed as a graph consisting of vertices (or nodes) connected by edges (or links). Erd\"{o}s and R\'{e}nyi, in 1960, defined an ER-network as $n$ vertices connected by $m$ edges, which are randomly chosen from totally possible $n(n-1)/2$ edges.  It has been proven for ER-network that the vertex degree distribution obeys a Poisson distribution\cite{Erdos-1960}. Yet, there are many real-life networks whose distributions are not Poisson distribution but heavily-tailed degree distribution, normally regarded as power-law distribution. In a network with power-law distribution, the most prominent feature is that  we often find a small number of highly connected vertices.  

In 1999, Barab\'{a}si and Albert \cite{Albert-1999-1} introduced a network in which the probability $P(k)$ of each vertex 
decays as a power-law
\begin{equation}\label{eq1}
 P(k)\sim k^{-\gamma}
\end{equation}
where $k$ is the degree of a vertex, and that network is called \emph{BA-network}. By observing many real networks, one has calculated the degree distribution of many real networks by equation (\ref{eq1}) and  discovered that their degree exponent $\gamma$ falls into an interval $(2,3]$. Based on equation (\ref{eq1}), Dorogovtsev \emph{et al.} \cite{Dorogovtsev-2002} gave the definition of cumulative degree distribution $P_{cum}(k)$ to explain the scale-free feature of deterministic networks as follows
\begin{equation}\label{eq2}
P_{cum}(k)=\sum_{k'\geq k}\frac{N(k', t)}{n_v(t)}\sim k^{1-\gamma}
\end{equation}
where $N(k',t)$ and $n_v(t)$ stand for the number of vertices with degree $k'$ and the network order at time step $t$, respectively; $k$ and $k'$ are positive integers of the discrete degree spectrum. Therefore, 
a network with scale-free behavior is illustrated by deducing that its degree distribution follows equation (\ref{eq1}) or its cumulative distribution obeys equation (\ref{eq2}), e.g. BA-model\cite{Albert-1999-1}and pseudofractal graphs\cite{Dorogovtsev-2002}. In this paper, we calculate the value of $\gamma$ for our networks by the cumulative degree distribution based on equation (\ref{eq2}) as we will shortly explain.

Although degree distribution describes some characteristics of the network, it is not enough for ones to analyze the network simply by degree distribution. Thus, other statistical indices have been developed to depict the nature of a network. Among them, the prevalent statistical indices are diameter and clustering coefficient of a network. If a network has a smaller diameter and a higher clustering coefficient, it can be referred to as a small-world network\cite{Watts-1998}. On the other hand, in our real life, the diameter of some networks generally increase exponentially with the number of verices\cite{Z-Z-2009}, while others usually grow logarithmically\cite{Zhang-2006,Lu-Guo-2012,Zhang-2009, Peng-2017}. Obviously, diameter plays a key role in determining the topological structures of complex networks. Hence, we here focus on the calculation process of diameter of all networks in our network space proposed shortly.

The clustering coefficient of a whole network can be obtained by the average of local clustering coefficient over all vertices in the network. The clustering coefficient of a vertex in network is defined as the ratio of the number of actually existing edges between all vertices adjacent to it and the number of all possible edges between them. According to the definition of clustering coefficient, the clustering coefficient of a vertex measures the network’s local edge density. The more densely interconnected the neighborhood of a vertex is, the higher the local clustering coefficient is. By using clustering coefficient, the observations show that some networks have a low clustering coefficient\cite{A-B-2002}, and however the remaining have a high clustering coefficient\cite{Zhang-2006,Lu-Guo-2012,Zhang-2009}. In other words, the clustering coefficient describes the phenomenon that vertices tend to create tightly knit groups characterized by a relatively high density of edges. So, we need to probe the clustering coefficient of all networks in network space.

The reminder of this paper is organized  by the following several sections. In Section II, we introduce three types of operations in detail, namely, rectangle operation, diamond operation, and triangle operation, and illustrate the process of generating a class of random scale-free networks $N(p,q,r,t)$ where probability parameters $p,q,r$ hold on $p+q+r=1$ with $0\leq p,q,r\leq 1$. After that, in Section III, we then discuss some topological structures of networks in our network space $\mathcal{N}(p,q,r,t)$, such as average degree, degree distribution, diameter, and clustering coefficient. Finally, for the outline of this paper, we have to draw a conclusion and bring some discussions for future work in the last section.

\section{Three types of Operations, Construction}

In this section, we will construct a class of random scale-free networks with tuning parameters which are generated from three different growth operations and then apply them to span a network space $\mathcal{N}(p,q,r,t)$. Here, the probability parameters $p,q,r$ hold $p+q+r=1$ with $0\leq p,q,r\leq1$, and $t$ stands for time step.  To this end, we firstly want to introduce three  graphic operations, named in this paper rectangle operation, diamond operation,  triangle operation, which are explained in more detail, as follows

\textbf{Rectangle operation.} The process can be divided into two steps, 

Step 1. For a given edge $uv$ with two vertices $u$ and $v$, we add an edge $xy$  with two endpoints $x$ and $y$.

Step 2. We connect the vertices $u$ with $x$ and $v$ with $y$. Then we obtain a cycle $C_{4}$.

Such a process is generally defined as rectangle operation since each edge generates a rectangle 
, see Fig.1.

\textbf{Diamond operation.} We have three steps,

Step 1. For a given edge $uv$ with two vertices $u$ and $v$, we create two vertices $x$ and $y$.

Step 2. We connect vertex $x$ with two endpoints $u$ and $v$ of the edge $uv$; at the same time, we link vertex $y$ with  two endpints $u$ and $v$ of the edge $uv$.

Step 3. We delete the edge $uv$. Then we also obtain a cycle $C_{4}$.

Such a process is often defined as diamond operation because each edge forms a diamond\cite{Ma-2019},  see Fig.1.

\textbf{Triangle operation.} We can divide this process into two steps, 

Step 1. For a given edge $uv$ with two vertices $u$ and $v$, we add a vertex $w$.

Step 2. We connect the vertex $w$ with $u$ and $v$, respectively.

Such a process is commonly defined as triangle operation since each edge generates a triangle, see Fig.1. 
\begin{figure}
\centering
\includegraphics[height=5cm]{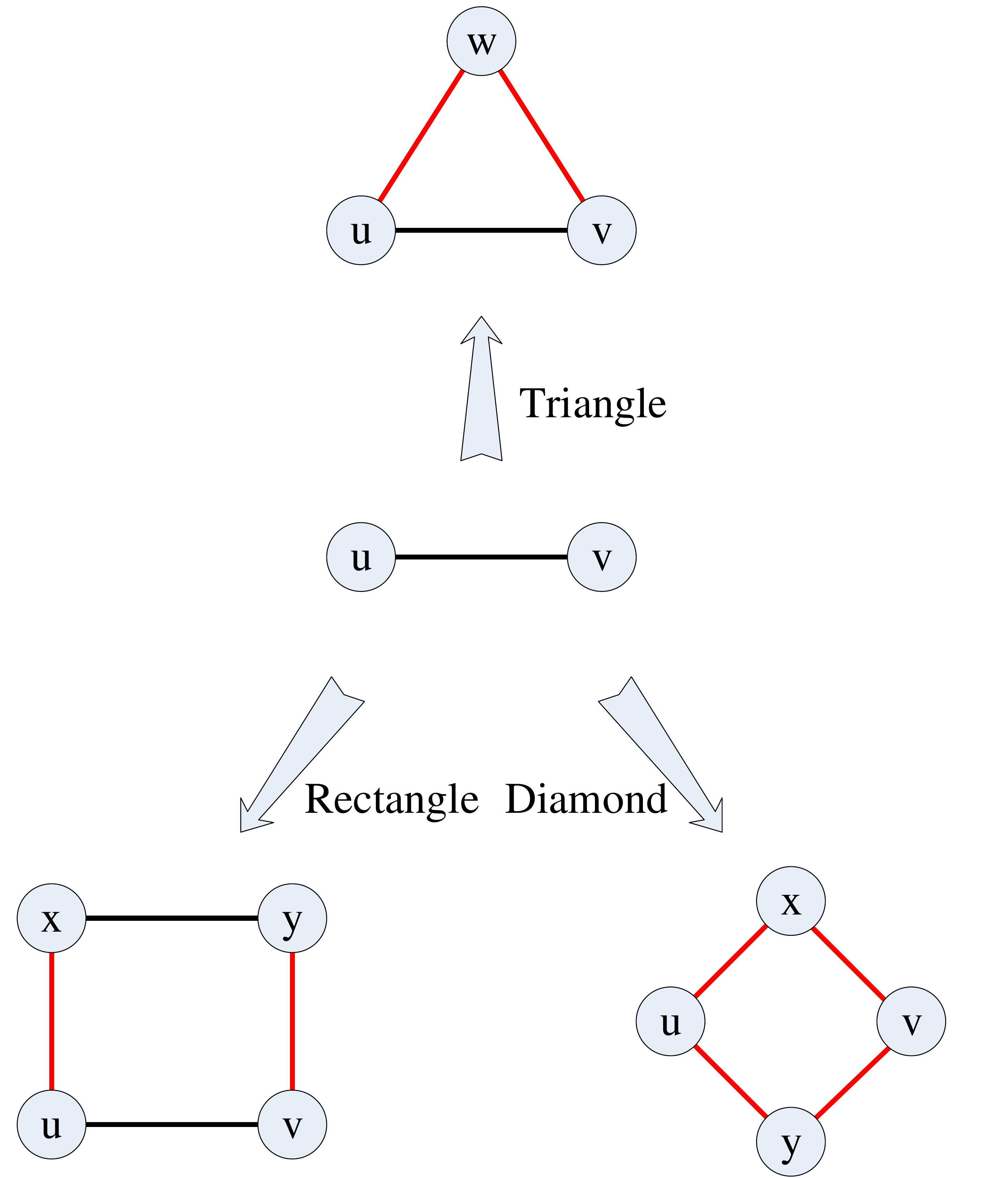}\\
{\small Fig.1. The diagram of rectangle operation, diamond operation as well as triangle operation.}
\end{figure}

According to the graph formed by the three types of operations, in this paper, we prefer to regard the first cycle $C_{4}$ as a rectangle, the second cycle $C_{4}$ as a diamond, and the third cycle $C_{3}$ as a triangle. In fact, from a theoretical perspective, it is not necessary to distinguish two cycles $C_{4}$ obtained from rectangle operation and diamond operation described above. In addition, because the clustering coefficient of $C_{4}$ is always equal to 0, while the clustering coefficient of $C_{3}$ is 1, it is obvious to find that $C_{4}$ and $C_{3}$ are completely different.  

Generally speaking, there exist two classical approaches to build the pre-existing networks. The first approach is to construct the network with some rules, such as ER-network \cite{Erdos-1960}, WS-network  \cite{Watts-1998}, BA-network \cite{Albert-1999-1}, NW-network \cite{Newman-W-1999}. The other is to generate an available network being in consistency with a designed degree sequence, e.g., configuration model, stochastic block model. 
Roughly speaking, networks generated by a designed degree sequence are more complicated than networks generated by rules in theory and practice. Since we have defined three types of operations, it is clear for the eye that we apply some rules to generate networks. The next task is to construct random scale-free networks $N(p,q,r,t)$ with three probability parameters $p,q,r$ holding on $p+q+r=1$ with $0\leq p,q,r\leq 1$ at each time step $t$. The degree of a vertex is the number of vertices of its neighbor set, represented by $k$. Let $|X|$ be the cardinality of set $X$. The concepts and symbols not given in this paper can be viewed in \cite{Bondy-2008}. 

\begin{algorithm}
  \SetKwData{Left}{left}
  \SetKwData{This}{this}
  \SetKwData{Up}{up}
  \SetKwFunction{Union}{Union}
  \SetKwFunction{FindCompress}{FindCompress}
  \SetKwInOut{Input}{Input}
  \SetKwInOut{Output}{Output}
  \caption{Construction-Pseudo code}
  \Input{The initial network $N(0)$ is an arbitrary graph with a fixed number of vertices,  $t$ stands for time step, probability parameters $p,q,r$ holding on $p+q+r=1$ with $0\leq p,q,r\leq 1$}
  \Output{The network $N(p,q,r,t)$}
  \BlankLine
  \For{$i\leftarrow 0$ \KwTo $t$}{
    {$a$ = random(0,1)}\;
    \While{edge to do exist}{
     \uIf{$a\in [0,p]$}{
       do rectangle operation;}
       \uElseIf{$a\in [p,p+q]$}{
       do diamond operation\;}
       \Else{
       do triangle operation\;}
       }
  }
  \label{algo_disjdecomp}
\end{algorithm}

\subsection{Constructions of random scale-free networks}
Taking useful advantage of three types of operations mentioned above, now let us turn our attention to constructing network space $\mathcal{N}(p,q,r,t)$. 

In fact, the initial network can be an arbitrary graph with a fixed number of vertices, and here, for the convenience of explanation, we use a cycle $C_{4}$ with four vertices as the initial network $N(0)$. Hence, one can easily get $N(p,q,r,t)$ from $N(p,q,r,t-1)$ for any time step $t\geq1$, i.e., utilizing the rectangle operation for each edge of network $N(p,q,r,t-1)$ with probability $p$ or employing the diamond operation for each edge of network $N(p,q,r,t-1)$ with probability $q$ or applying the triangle operation for each edge of network $N(p,q,r,t-1)$ with probability $r$, shown in Fig.2. The algorithm 1 illustrates the construction of generating our networks, where $a$ is a random number which falls into an interval [0,1]. The choice of an edge is uniformly random. It should be specially noted that, for each edge, three types of operations are mutually exclusive. Put it another way, if the rectangle operation is performed on a certain edge, the diamond operation and the triangle operation cannot be performed on this edge. As previously discussed, after $t$ time steps, there will be $(4\times(4p+4q+3r)^{t})^{3}=(4\times(4-r)^{t})^{3}$ members (including isomorphic networks) in our network space $\mathcal{N}(p,q,r,t)$. By adjusting the value of probability parameters $p,q,r$, the resulting networks in our space will become much richer and colorful. In particular, our network space $\mathcal{N}(p,q,r,t)$ contains a subspace $\mathcal{N}(p,q,0,t)$ without triangle operation and a subnetwork $N(0,0,r,t)$ with triangle operation. The topological structures of the networks $\mathcal{N}(p,q,0,t)$ and $N(0,0,r,t)$ have been thoroughly investigated in \cite{Ma-2019,W-Y-2016}. Except for that, our network space contains a large variety of random scale-free networks, to which we will pay more attention. In the following, we will discuss the topological properties of random scale-free networks in our space.


\begin{figure}
\centering
\includegraphics[height=5cm]{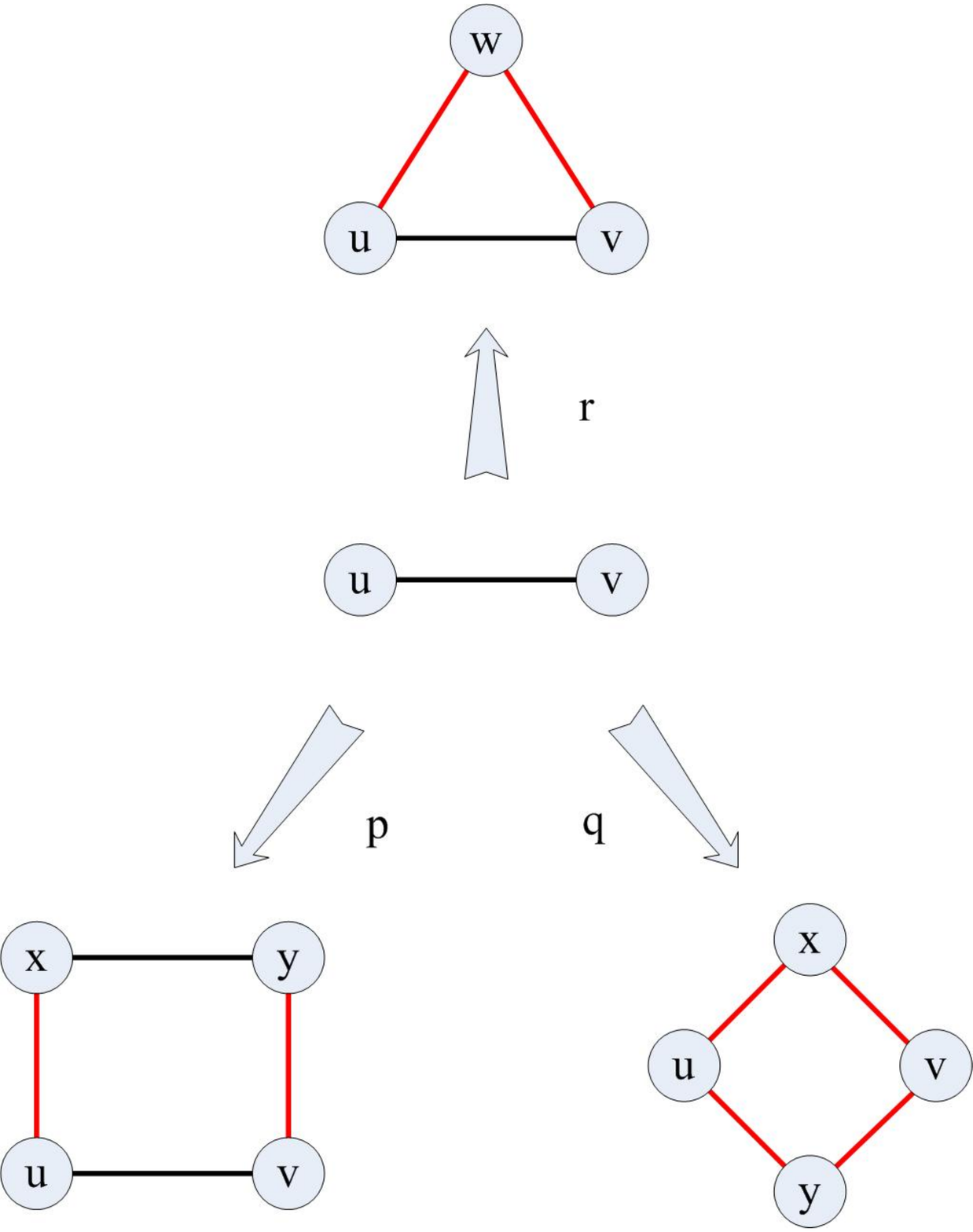}\\
{\small Fig.2. The diagram of operation implemented on an edge.  The choice of edge is uniformly random. For each edge, only one type of operation can be selected from our operations. In the process of constructing $N(p,q,r,t)$ from $N(p,q,r,t-1)$ ($t\geq1$), $p$ represents the probability of applying rectangle operation to each edge of $N(p,q,r,t-1)$; $q$ represents the probability of applying diamond operation to each edge of $N(p,q,r,t-1)$; $r$ represents the probability of applying triangle operation to each edge of $N(p,q,r,t-1)$; and $p+q+r=1, 0\leq p,q,r\leq 1$.}
\end{figure}

\section{Topological properties of random scale-free networks}

In order to better understand the properties of our network $N(p,q,r,t)$, we discuss some related quantities that determine topological structures, such as average degree, degree distribution, diameter, and clustering coefficient. What we present in this section will consist of the following subsections.

\subsection{Average degree}

Firstly, we calculate two basic quantities, namely, the number of vertices  $|V(p,q,r,t)|$ and edges number $|E(p,q,r,t)|$ in $N(p,q,r,t)$, also called network order and size, respectively. According to three types of operations, the order and size have the relationship in the below equation
\begin{align*}
|V(p,q,r,t)|&=(2p+2q+r)|E(p,q,r, t-1)|+|V(p,q,r, t-1)|\\
|E(p,q,r,t)|&=(4p+4q)|E(p,q,r, t-1)|+3r|E(p,q,r, t-1)|
\end{align*}
It is not difficult to obtain the order and size of our networks $N(p,q,r,t)$ as follows
\begin{align*}
|V(p,q,r,t)|&=4\times(2p+2q+r)\frac{(4p+4q+3r)^{t}-1}{4p+4q+3r-1}+4\\ &=4+4\times(2-r)\frac{(4-r)^{t}-1}{3-r}\\
|E(p,q,r,t)|&=4\times(4p+4q+3r)^{t}=4\times(4-r)^{t}
\end{align*}
Clearly, the order and size of member $N(p,q,r,t)$ in network space $\mathcal{N}(p,q,r,t)$ are adjustable values by varying time step $t$ and probability parameter $r$. According to the definition of average degree\cite{Albert-1999-1}, the average degree $\langle k\rangle$ of our networks can be calculated as below
\begin{align*}
\langle k\rangle=\frac{2|E(p,q,r,t)|}{|V(p,q,r,t)|}=\frac{2\times 4\times (4-r)^{t}}{4+4\times(2-r)\frac{(4-r)^{t}-1}{3-r}}
\end{align*}

One can easily discover that $\langle k\rangle$ is associated with $t$ and $r$.  In the limit of time step $t$,
\begin{align*}
\langle k\rangle=\frac{2\times 4\times (4-r)^{t}}{4+4\times(2-r)\frac{(4-r)^{t}-1}{3-r}}\approx\frac{6-2r}{2-r}
\end{align*}
Obviously, it is not hard to reveal that $\langle k\rangle$ is independent of $p,q$, and completely determined by $r$.

In complex networks study, a network is sparse if $|E(p,q,r,t)|\ll \frac{|V(p,q,r,t)|\times(|V(p,q,r,t)|-1)}{2}$, i.e., the number of all edges is much smaller than the maximum number of edges. 

Since $r\in[0,1]$, at $r=0$, the average degree $\langle k\rangle$ of $N(p,q,r,t)$, $\langle k\rangle\approx\frac{6-2r}{2-r}=\frac{6-2\times 0}{2-0}$, is small and approximately equal to $3$.  It should be mentioned that the network with the same average degree has been discussed in\cite{Ma-2019}. For a large enough $t$, we can state that the resulting networks $N(p,q,r,t)$ are sparse networks as many real-world networks whose vertices have many fewer connections than is possible. At $r=1$, the average degree $\langle k\rangle$ of $N(p,q,r,t)$, $\langle k\rangle\approx\frac{6-2r}{2-r}=\frac{6-2\times 1}{2-1}$, is close to $4$. It is interesting to note that the identical average degree has been observed analytically in \cite{Z-Z-2009}, Pseudofractal graphs \cite{Dorogovtsev-2002}, recursive graphs \cite{Comellas-2004}, and Apollonian networks\cite{zhang-2006}. Our finding shows that the resulting networks $N(p,q,r,t)$ are sparse networks when the time step $t$ tends to infinity.

\begin{figure}
\centering
\includegraphics[height=5cm]{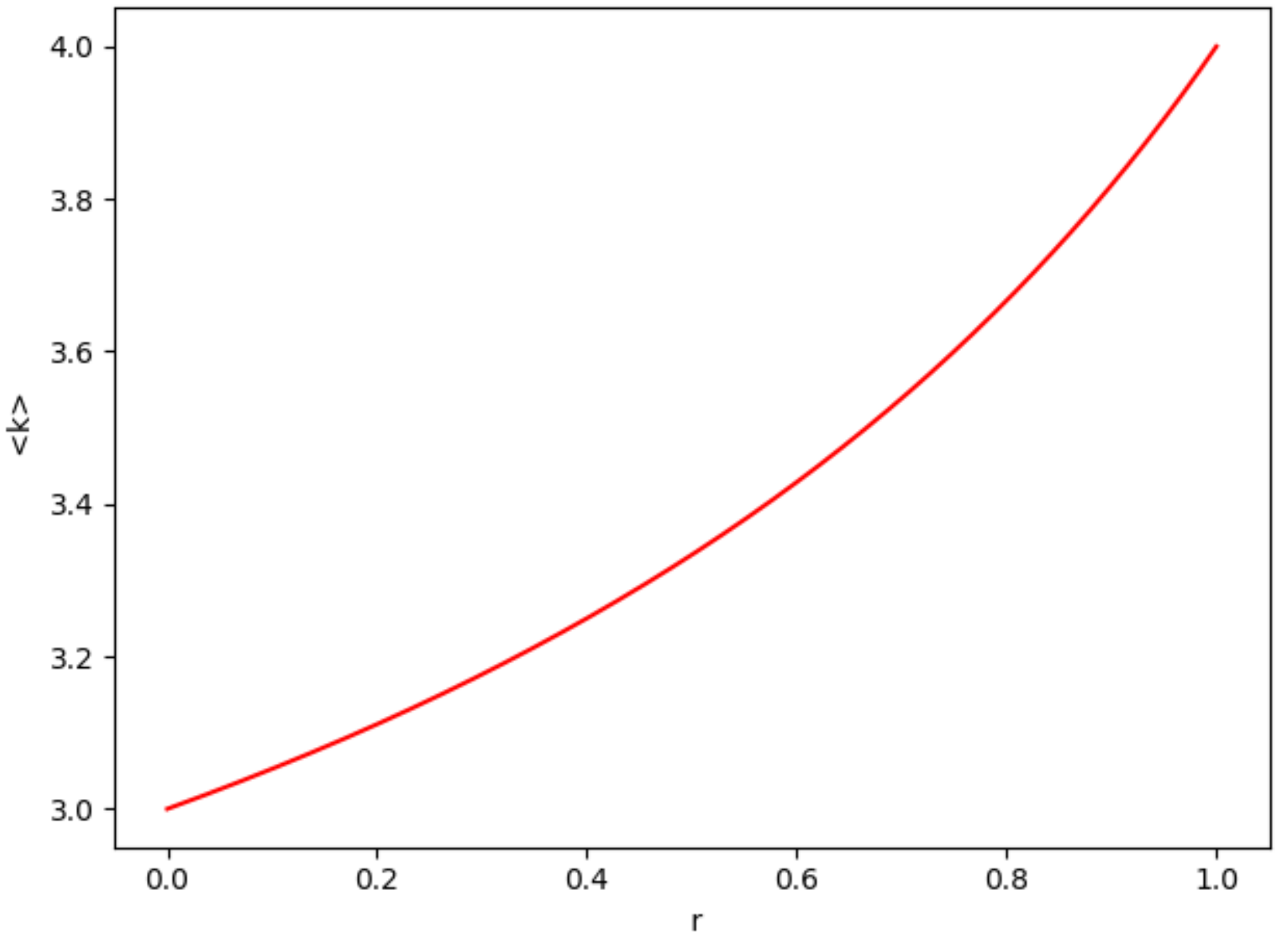}\\
{\small Fig.3. The diagram of average degree with the probability parameter $r$,  $0\leq r\leq 1$, of network space $\mathcal{N}(p,q,r,t)$.  It is clear for the eye that the curve illustrates the value of average degree changing with probability parameter $r$.}
\end{figure}

Based on our above analysis and Fig.3, we have the following proposition.

\textbf{Proposition 1} For any member $N(p,q,r,t)$ of network space $\mathcal{N}(p,q,r,t)$, the average degree $\langle k\rangle$ must hold the following inequality
\begin{equation}\label{average}
3\leq\langle k\rangle\approx\frac{6-2r}{2-r}\leq4.
\end{equation}

As is clear from equation (\ref{average}), it is worth noting that each $r$ corresponds to a unique $\langle k\rangle$.  It goes without saying that $\langle k\rangle$ monotonically increases with $r$. The equation (\ref{average}) also indicates that the value of average degree of our networks with mix graphic operations will fall into a unique interval $[3,4]$. Besides, the average degree of each member in network space has nothing to do with initial network, the number of all vertices, and the number of all edges. One can find that the number of all edges of the network in our space is only a tiny fraction of the expected number of edges for a complete graph of the same number of vertices. 
Actually, $r$ determines the average degree and then average degree determines the sparsity. So, we assert that each network in our space $\mathcal{N}(p,q,r,t)$ is always sparse.

\subsection{Degree distribution}

Degree distribution is one of most fundamental and important topological structures of a network, which is the standard for judging whether the network is scale-free. The following analysis shows that our networks are scale-free.

Now we analyze the degree distribution of our networks $N(p,q,r,t)$. When a new vertex $i$ is connected to the network at a certain time step $t_{i}$($t_{i}\geq 1$), it is not difficult to find that the vertex $i$ has a degree of 2 no matter which operation is taken. We denote by $k_{i}(t)$ the degree of vertex $i$ at time step $t$. From the construction process of network, the degree $k_i(t)$ evolves with time as $k_i(t) = 2k_i(t-1)=2^{t+1-t_i}$. The degree of vertex $i$ is increased by a factor 2 at each time step.

Given the process of constructing our networks, we can see that the degree spectrum of our networks in our space is a series of discrete real values. In order to calculate the power-law exponent of the degree distribution of our networks, we take full advantage of the method proposed by Dorogovtsev\cite{Dorogovtsev-2002}, as shown in equation (\ref{eq2})
\begin{equation*}\label{cum-11}
P_{cum}(k)=\frac{|V(p,q,r,t_{i})|}{|V(p,q,r,t)|}=\frac{4+
4\times(2-r)\frac{(4-r)^{t_i}-1}{3-r}}{4+
4\times(2-r)\frac{(4-r)^{t}-1}{3-r}}
\end{equation*}
where $P_{cum}(k)$ is the probability that the degree of a vertex in our network is greater than $k_{i}(t)$.

We omit the constant for a very large $t$, and then get the formula as below.
\begin{equation*}\label{cum-111}
P_{cum}(k)=\frac{4+
4\times(2-r)\frac{(4-r)^{t_i}-1}{3-r}}{4+
4\times(2-r)\frac{(4-r)^{t}-1}{3-r}}\approx(4-r)^{t_{i}-t}
\end{equation*}
Together with $k_{i}(t)=2^{t-t_{i}+1}$, we have $t_{i}=t+1-\frac{\ln k}{\ln 2}$. Then, we plug $t_i$ into the above expression and obtain
\begin{equation*}\label{cum-22}
P_{cum}(k)\sim k^{1-\gamma}
\end{equation*}
where $\gamma=1+\frac{\ln(4-r)}{\ln2}$.

Therefore, the cumulative degree distribution $P_{cum}(k)$ of network $N(p,q,r,t)$ follows a power-law form  with the degree exponent $\gamma=1+\frac{\ln(4-r)}{\ln2}$. It is evident that $\gamma$ is independent of $p,q$, and is related to $r$.

Thanks to $r\in[0,1]$, at $r=0$, we can find that the network $N(p,q,r,t)$ follows a power-law with the degree exponent $\gamma=1+\frac{\ln(4-0)}{\ln2}=3$. Note that the same degree exponent has been obtained in the deterministic networks \cite{Z-Z-2009}, classical BA networks \cite{A-L-2003} and typical random Sierpinksi networks \cite{Z-Z-2008}, and so on. At $r=1$, one can discover that the degree exponent $\gamma$ of the network $N(p,q,r,t)$ obeys a power-law with $\gamma=1+\frac{\ln(4-1)}{\ln2}=1+\frac{\ln3}{\ln2}$. Apparently, $2<\gamma=1+\frac{\ln3}{\ln2}\leq3$, it indicates that the network $N(p,q,r,t)$ is scale-free. It is noticeable that the same degree exponent has been obtained in Pseudofractal graphs \cite{Dorogovtsev-2002}, recursive graphs  \cite{Comellas-2004}, Apollonian networks \cite{zhang-2006}, and deterministic scale-free networks\cite{A-L-2001}, etc.

\begin{figure}
\centering
\includegraphics[height=5cm]{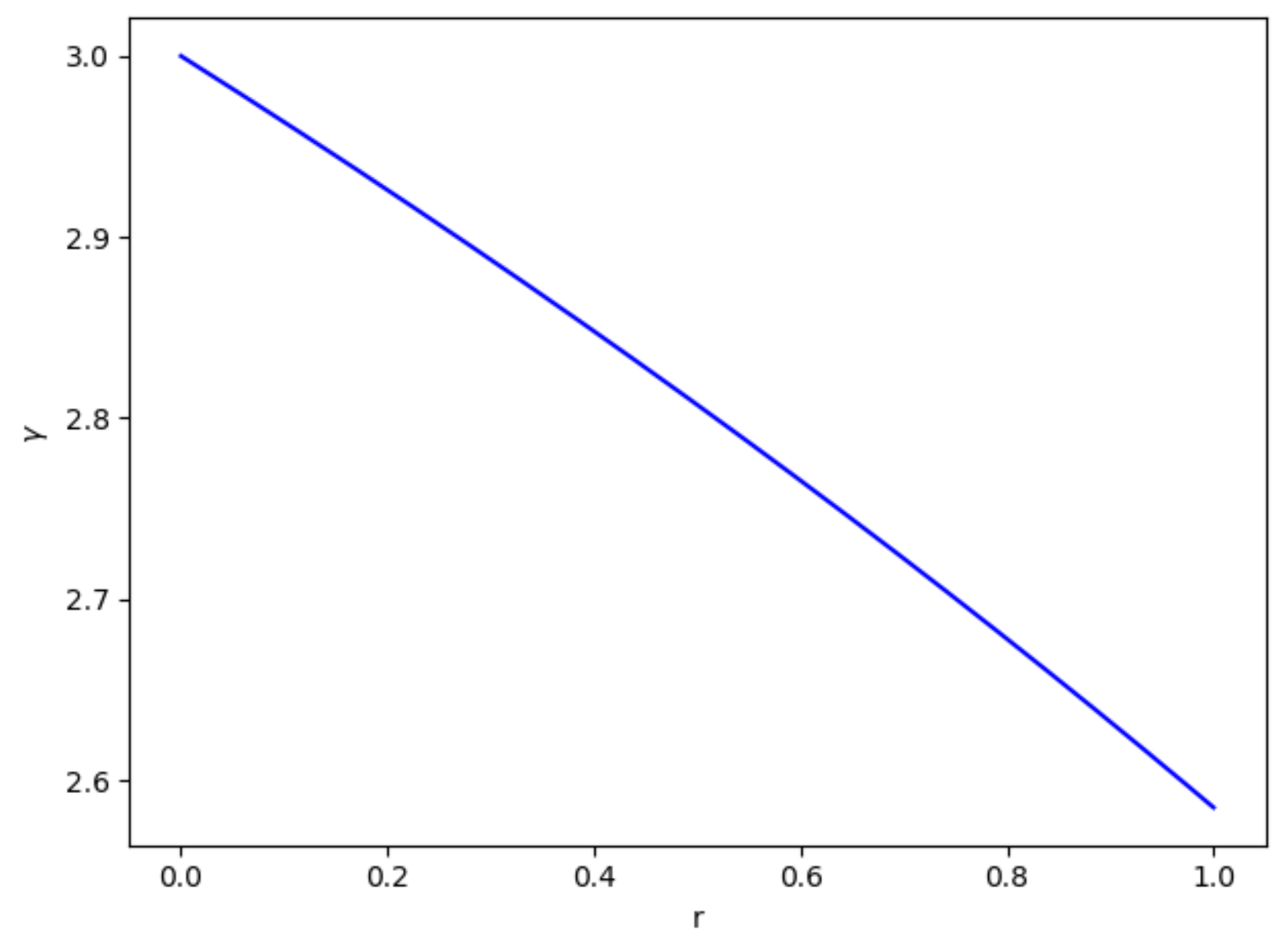}\\
{\small Fig.4. The diagram of degree exponent with probability parameter $r$, $0\leq r\leq 1$, of network space $\mathcal{N}(p,q,r,t)$. Obviously, the curve indicates the value of degree exponent varies with the probability parameter $r$.}
\end{figure}

Combining with previous analysis and Fig.4, it is not hard to obtain the following proposition.

\textbf{Proposition 2} For any member $N(p,q,r,t)$ of network space $\mathcal{N}(p,q,r,t)$, the power-law exponent of  $N(p,q,r,t)$ must satisfy the following inequality
\begin{equation}\label{degree}
2<1+\frac{\ln 3}{\ln 2} \leq \gamma=1+\frac{\ln(4-r)}{\ln2} \leq 3.
\end{equation}

From equation (\ref{degree}), we note that each $r$ corresponds to a unique $\gamma$. It can be said with certainty that $\gamma$ decreases monotonically with $r$ as we can see from the equation (\ref{degree}) and Fig.4. This result shows that the value of power-law exponent of our networks in our space will fall into a restricted interval.
In addition, it is apparent that each member in network space $\mathcal{N}(p,q,r,t)$ is a scale-free network due to degree distribution obeys power-law distribution with degree exponent in (2,3]. Consequently we can induce  the fact that all members in our network space $\mathcal{N}(p,q,r,t)$ are scale-free. 


\subsection{Diameter}

The small-world concept depicts the fact that there is a relatively smaller distance $d_{uv}$ between any pair of vertices $u$ and $v$ in a network. The distance between two vertices is the smallest number of edges to get from $u$ to $v$. The longest shortest path between all pairs of vertices is called diameter.  Diameter is itself a feature of network structure and can be applied to characterizing communication delay over a network. In general, the larger diameter is, the lower communication efficiency is. The diameter of our network is denoted as $D(p,q,r,t)$. Here, we will introduce the main idea of analysis.

Before continuing, let us concentrate on three special cases, which help us to deduce iterative expressions for diameters of among  networks  $N(1,0,0,t)$, $N(0,1,0,t)$, and $N(0,0,1,t)$, respectively. Now we discuss the calculation process of diameter in three cases.

(a) In the procedure of producing $N(1,0,0,t)$ from $N(1,0,0,t-1)$, one can notice that the diameter $D(1,0,0,t)$ is deduced by $D(1,0,0,t-1)$. The derivation is available for all diameters of network $N(1,0,0,t-1)$. Consequently, we may find a recursive expression between $D(1,0,0,t)$ and $D(1,0,0,t-1)$, $D(1,0,0,t)=D(1,0,0,t-1)+2$. Together with known initial condition $D(1,0,0,0)=2$, we obtain a closed-form formula of diameter $D(1,0,0,t)$, i.e., $D(1,0,0,t)=2(t+1)$ for all $t\geq 2$. From another point of view, there is a fact that $\ln|V(1,0,0,t)|\sim\ln4^{t+1}=(t+1)\ln4$. Then, we discover a connection between diameter $D(1,0,0,t)$ and order $|V(1,0,0,t)|$ under the approximate relationship $D(1,0,0,t)\sim\ln|V(1,0,0,t)|$. Hence, as $t$ approaches infinity, the diameter $D(1,0,0,t)$ scales logarithmically with the network order. This cases can be easily found in many real networks\cite{Watts-1998}, which implies that the network $N(1,0,0,t)$ is similar to these real networks. 

(b) Consider the process of generating $N(0,1,0,t)$ from $N(0,1,0,t-1)$, there is no deny that the diameter $D(0,1,0,t)$ is derived from $D(0,1,0,t-1)$. So, we can find an iterative relationship between diameter $D(0,1,0,t)$ and $D(0,1,0,t-1)$, i.e., $D(0,1,0,t)=2D(0,1,0,t-1)$. 
Combined with initial condition $D(0,1,0,0)=2$, we can infer a solution of diameter $D(0,1,0,t)$, i.e., $D(0,1,0,t)=2^{t+1}$ for all $t\geq 2$. Therefore, we may find that diameter $D(0,1,0,t)$ is not approximately equal to $\ln|V(0,1,0,t)|$ but a square value of order of network $N(0,1,0,t)$, directly showing $D(0,1,0,t)$ is large-scale. Obviously, when the order of network is large, diameter $D(0,1,0,t)$ increases exponentially.

(c) By analogy with the diameter $D(1,0,0,t)$, we can deduce the diameter $D(0,0,1,t)$ by $D(0,0,1,t-1)$.  Then, we also reveal a recursive connection between $D(0,0,1,t)$ and $D(0,0,1,t-1)$, $D(0,0,1,t)=D(0,0,1,t-1)+2$. Together with initial condition $D(0,0,1,0)=2$, we can also get a closed-form formula of diameter $D(0,0,1,t)$, i.e., $D(0,0,1,t)=2(t+1)$ for any $t\geq 2$. From another angle, there exists a result that $\ln|V(0,0,1,t)|\sim\ln4^{t+1}=(t+1)\ln4$.  Then we find a relationship between diameter and order, namely,  $D(0,0,1,t)\sim\ln|V(0,0,1,t)|$. So, as $t$ goes to infinity, the diameter $D(0,0,1,t)$ grows logarithmically with network order. It goes without saying that $D(1,0,0,t)$ is equal to $D(0,0,1,t)$. 

So far, we have studied three concrete examples in practice, while the discussion about diameters of other members in our network space $\mathcal{N}(p,q,r,t)$ have not been explored thoroughly. One of the most important reasons is to introduce probability parameters $p,q,r$ into the network generation. 

On the basis of three deterministic cases and our analysis, we try to calculate an analytical solution to diameter of our network. Different from the previous calculation, we need to consider several contributions to the diameter change by implementing rectangle, diamond and triangle operations in the process of obtaining network $N(p,q,r,t)$ from  $N(p,q,r,t-1)$. We randomly choose an arbitrary path of length equivalent to $D(p,q,r,t-1)$, denoted as $u_{1}u_{2},...,u_{D(p,q,r, t-1)+1}$. Such choice is reasonable because we assume that each edge can only choose one from three operations of rectangle, diamond and triangle operation. We can only do one operation on an edge at a certain time $t$. For an edge $uv$, if rectangle operation is selected, diamond and triangle operation cannot be selected for the edge $uv$.

Case 1. Assume we apply rectangle operation on edges both  $u_1u_2$ and $u_{D(p,q,r,t-1)}u_{D(p,q,r,t-1)+1}$, at the same time, then we have  $D(p,q,r,t)_1=(p+r)D(p,q,r,t-1)+2(1-q)+2qD(p,q,r,t -1)$. 

Case 2. Provided that only one of the two edges $u_1u_2$ and $u_{D(p,q,r,t-1)} u_{D( p,q,r,t-1)+1}$ is performed the rectangle operation, then $D(p,q,r,t)_2=(p+r)D(p,q,r,t-1)+1-q+2qD(p,q,r,t -1)$. 

Case 3. Suppose that neither edge $u_1u_2$ nor edge $u_{D(p,q,r,t-1)}u_{D(p,q,r,t-1)+1}$ are not conducted the rectangle operation, and so we obtain $D(p,q,r,t)_3=(p+r)D(p,q,r,t -1)+2q D(p,q,r,t-1)$.

Obviously, one can discover that $D(p,q,r,t)_1\geq D(p,q,r,t)_2\geq D(p,q,r,t)_3$ according to the above three cases. Hence, we in theory compute the maximum value of $D(p,q,r,t)$ of network $N(p,q,r,t)$. From what has been mentioned above, we may easily find an iterative expression as below  
\begin{align*}\label{dia}
D(p,q,r,t)&=(p+r)(D(p,q,r,t-1)+2)+2q D(p,q,r,t-1)\\
&=(1+q)D(p,q,r,t-1)+2-2q
\end{align*}
Together with initial condition, we have
\begin{equation*}\label{dia1}
D(p,q,r,t)=(1+q)^{t}M+(2-2q)t
\end{equation*}
where $M=D(p,q,r,0)$. Therefore, it is clear that $D(p,q,r,t)$ is independent of $p,r$ and is relevant with $q$ and $t$.

Because $q\in[0,1]$, at $q=0$, we can compute the diameter $D(p,q,r,t)$ of  $N(p,q,r,t)$, $D(p,q,r,t)=(1+0)^{t}M+2t$, then we get $D(p,q,r,t)=2(t+1)$ which linearly increases with time for all $t\geq 1$. Note that the same diameter has been obtained in the deterministic network\cite{zhang-2006}. It should be mentioned that the diameter $D(p,q,r,t)$ of network in our space that does not contain the diamond operation grows logarithmically with network order. This phenomena can be easily found in a large quantity of real networks, which indicates that network $N(p,q,r,t)$ is like such real networks with small diameter.

At $q=1$, we obtain $D(p,q,r,t)=(1+1)^{t}\times D(p,q,r,t-1)$. We can easily find $D(p,q,r,t)=2^{t+1}$ for all $t\geq 2$. It is noticeable that the diameter $D(p,q,r,t)$ increases exponentially, and is not linear with network order but a square root value of network order.


\begin{figure}
\centering
\includegraphics[height=5cm]{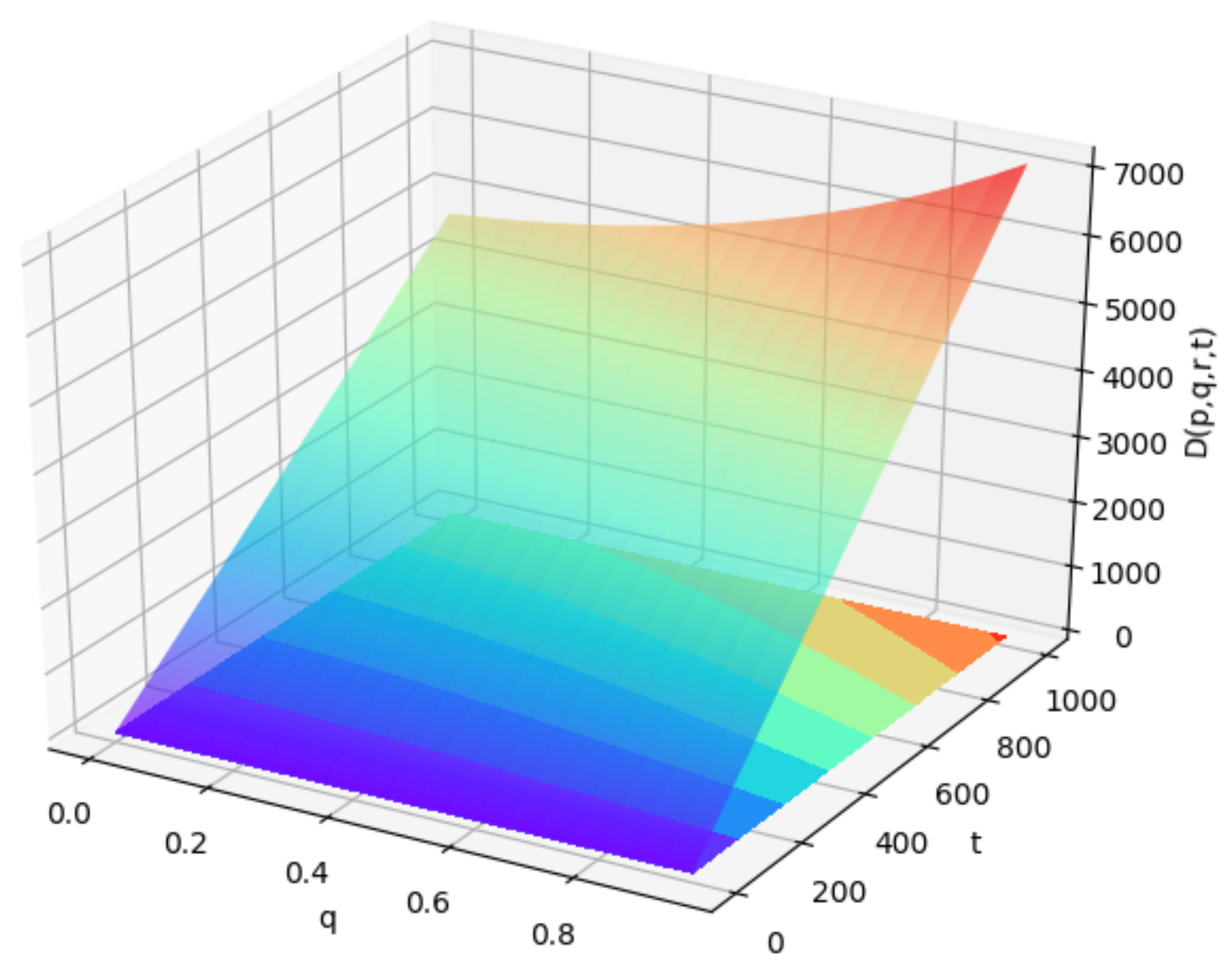}\\
{\small Fig.5. The diagram of diameter with constraints,  $0\leq t\leq 1000$ and $0\leq q \leq 1$, of network space $\mathcal{N}(p,q,r,t)$.  It also demonstrates that the curve represents the value of diameter mainly depends on time $t$ and probability parameter $q$.}
\end{figure}

On account of aforementioned analysis and Fig.5, one can get the following proposition.

\textbf{Proposition 3} For any member $N(p,q,r,t)$ of network space $\mathcal{N}(p,q,r,t)$, the diameter of  $N(p,q,r,t)$ must meet the following inequality
\begin{equation}\label{diameter}
2(t+1)\leq D(p,q,r,t)=(1+q)^{t}M+(2-2q)t \leq 2^{t+1}.
\end{equation}

From equation (\ref{diameter}), it is worth noticing that the diameter is mainly affected by $q$ and $t$. The diameter increases monotonically with $t$ and $q$. This result shows that the value of diameter of the network in our space will fall into a restricted interval. 
For a network in our space, the diameter of the network may abruptly transform from a small value to a high value if we do the diamond operation for each edge of the network. Together with our analysis, we discover that the diameters of these members in our space that do not contain the diamond operation grow linearly with time $t$ or logarithmically with network order, otherwise, the diameters increase exponentially. In addition, to understand the network's behavior, we need to became familiar with the clustering coefficient, which is the main part in coming subsections.





\subsection{Clustering coefficient}

Clustering is another vital property of a network, which provides measure of local structure within the network. The most immediate measure of clustering is the clustering coefficient $c_i$ for every vertex $i$. By definition, clustering coefficient of a vertex $i$ is the ratio of the total number $E_i$ of edges that actually exist between all $k_i$ its nearest neighbors and the number of $k_{i}(k_{i}-1)/2$ of all possible edges between them, i.e., $c_{i}=2E_{i}/[k_{i}(k_{i}-1)]$. The clustering coefficient $\langle c\rangle$ of the entire network is average of all vertex $c_{i}$'s.

Now we will compute the clustering coefficient of every vertex and their average value. As shown in previous researches\cite{Newman-W-1999}, most networks are highly transitive or clustered, i.e., a friend may be two friends individually, who may then become acquainted with one another through their common friend, and so end up friends themselves. To better simulate the actual network, a subspace with non-clustering have been investigated in our another paper\cite{Ma-2019}. Here, our next task is to research the clustering coefficient of networks by varying the value of $r$ as we will see in the following.



Armed with previous discussion,  we find that there are two factors have an effect on the average clustering coefficient of the network $N(p,q,r,t)$. The first is the value of $r$ and the second is that the total number of times of performing the triangle operation at time $t$.


Hence, we can divide the process of average clustering coefficient into 2 steps. Firstly, we find the clustering coefficient for each vertex in $N(p,q,r,t)$, derive a closed formula for the clustering coefficient $\langle c\rangle$, and then list in Table-I 
\begin{center}
Table-I the clustering coefficient of $c_{i}$ of vertices degree $k_{i}$
\vskip 0.2cm
\begin{tabular}{c|c|c|c|c|c|c|c}
\hline
$k_i$& $2$&$2^{2}$ & $2^{3}$ & $\cdots$ & $2^{t-1}$ &$2^{t}$ & $2^{t+1}$ \\
\hline
$c_i$ & $1$ & $\frac{1}{3}$& $\frac{3}{14}$& $\cdots$ & $\frac{ (2^{t-1}-2)}{2^{t-2}(2^{t-1}-1)}$ &$\frac{2^{t}-2}{2^{t-1}(2^{t}-1)}$ & $\frac{2^{t+1}-2}{2^{t}(2^{t+1}-1)}$\\
\hline
\end{tabular}
\end{center}

Secondly, we calculate the proportion of the vertices with a clustering coefficient of $c_{i}$ in network $N(p,q,r,t)$. In order to obtain the clustering coefficient $\langle c\rangle$ of the entire network $N(p,q,r,t)$, it is necessary for us to give the degree distribution spectrum, namely, the probability $p_{i}(p_{i}=\frac{n_{k_{i}(t)}}{|V(p,q,r,t)|})$ of vertices degree $k_i$ show in Table-II
\begin{center}
Table-II the degree spectrum of degree $k_{i}$
\vskip 0.2cm
\begin{tabular}{c|c|c|c|c|c|c|c}
\hline
$k_i$& $2$&$2^{2}$ & $2^{3}$ & $\cdots$ & $2^{t-1}$ &$2^{t}$ & $2^{t+1}$ \\
\hline
$p_i$ & $\frac{3-r}{4-r}$ & $\frac{3-r}{(4-r)^{2}}$& $\frac{3-r}{(4-r)^{3}}$& $\cdots$ & $\frac{3-r}{(4-r)^{t-1}}$ &$\frac{3-r}{(4-r)^{t}}$ & $\frac{3-r}{(4-r)^{t}}$\\
\hline
\end{tabular}
\end{center}

On the basis of above discussions, it is not difficult to obtain the clustering coefficient $\langle c\rangle$ of the entire network is
\begin{equation}\label{cluster1}
\langle c\rangle =r\sum p_{i}c_{i}
\end{equation}

Therefore, it is clear for the eye that $\langle c\rangle$ is independent of $p,q$ and is related to $r$. Owing to $r\in[0,1]$, for first case with $r=0$, the clustering coefficient of every vertex in $N(p,q,0,t)$ is zero. Hence, the average value of clustering coefficient in $N(p,q,0,t)$ is always equal to 0. So, we say that the clustering coefficient of whole network is zero.

For second case with $r=1$, it is noticeable that the clustering coefficient is approximately to 0.7566. Apparently, when the parameters $r$ is larger, the clustering coefficient is higher.

As mentioned previously, it is not difficult to derive the following proposition.

\textbf{Proposition 4} For any member $N(p,q,r,t)$ of network space $\mathcal{N}(p,q,r,t)$, the clustering coefficient must hold the following inequality
\begin{equation}\label{clusering1}
0 \leq \langle c\rangle=r\sum p_{i}c_{i}\leq 0.7566
\end{equation}

From equation (\ref{clusering1}), it has to be noticed that clustering coefficient is related to $r$ and monotonically increases with $r$. This result shows that the value of clustering coefficient of the network in our space will fall into a restricted range, where the lower bound and the upper bound are the minimum and maximum values of its constituents, respectively. Hence, the clustering coefficient of all members in network space $\mathcal{N}(p,q,r,t)$ range from 0 to 0.7566. 

\section{Conclusion and Discussions}

In this paper, we have introduced three types of operations and presented the process of generating random scale-free networks, which constitute the network space $\mathcal{N}(p,q,r,t)$. And then, we have discussed their topological properties such as, average degree, degree distribution, diameter, and clustering coefficient. Above all, we have computed the average degree of each member in our network space and discovered that all members are sparse. Secondly, we have proven that the degree distribution obeys the power-law distribution, which means that all networks are always scale-free. 

It is clear that the resulting network space contains many members by adjusting the value of probability parameter $p,q,r$. For example, by choosing different $p,q,r$ at different time steps, we can find that the resulting networks in our network space are more colorful. Especially, the diameter may abruptly transform from zero to high value. Accomplishment notwithstanding, research on complex networks is far from enough and requires long-term long-sustainable endeavor. New discoveries, developments, enhancements, and improvements are still needed. In the future, we will devote more efforts to investigating complex networks in order to better help people apply it to explain some phenomena in real life.

\begin{acknowledgments}
We are grateful to the anonymous referees for their valuable and helpful comments which lead to the improvement of this paper. This research was supported by the National Natural Science Foundation of China under grants  No. 61662066.
\end{acknowledgments}

\end{document}